\documentclass[sigconf]{acmart}
\usepackage{amsmath} 
\usepackage{array}
\usepackage{amsfonts}
\usepackage{dsfont}
\usepackage{float}
\usepackage[utf8]{inputenc}
\DeclareUnicodeCharacter{03B1}{$\alpha$}

\usepackage{tabularx} 
\usepackage{booktabs} 
\AtBeginDocument{%
  \providecommand\BibTeX{{%
    \normalfont B\kern-0.5em{\scshape i\kern-0.25em b}\kern-0.8em\TeX}}}

\copyrightyear{2026}
\acmYear{2026}
\setcopyright{cc}
\setcctype{by}
\acmConference[AHs 2026]{The Augmented Humans International Conference 2026}{March 16--19, 2026}{Okinawa, Japan}
\acmBooktitle{The Augmented Humans International Conference 2026 (AHs 2026), March 16--19, 2026, Okinawa, Japan}
\acmDOI{10.1145/3795011.3795035}
\acmISBN{979-8-4007-2351-3/2026/03}

\begin{document}

\title{OpticalAging: Real-time Presbyopia Simulation for Inclusive Design via Tunable Lenses}

\author{Qing Zhang}
\orcid{0000-0002-0622-932X}
\affiliation{%
  \institution{The University of Tokyo}
  \city{Tokyo}
  \country{Japan}}
\email{qzkiyoshi@gmail.com}

\author{Zixiong Su}
\orcid{0000-0001-6048-3268}
\affiliation{%
  \institution{The University of Tokyo}
  \city{Tokyo}
  \country{Japan}}
\email{soshiyuu@gmail.com}

\author{Yoshihito Kondoh}
\orcid{0009-0000-2776-3620}
\affiliation{%
  \institution{ViXion Inc.}
  \city{Tokyo}
  \country{Japan}
}
\email{yoshihito.kondoh@vixion.jp}

\author{Kazunori Asada}
\affiliation{%
  \institution{ViXion Inc.}
  \city{Tokyo}
  \country{Japan}}
\email{kazunori@k-asada.com}

\author{Thad Starner}
\orcid{0000-0001-8442-7842}
\affiliation{%
  \institution{Georgia Institute of Technology}
  \city{Atlanta}
  \country{USA}}
\email{thad@gatech.edu}

\author{Kai Kunze}
\orcid{0000-0003-2294-3774}
\affiliation{%
  \institution{Keio University}
  \city{Yokohama}
  \country{Japan}}
\email{kai@kmd.keio.ac.jp}

\author{Yuta Itoh}
\orcid{0000-0002-5901-797X}
\affiliation{
    \institution{Institute of Science Tokyo}
    \city{Yokohama}
    \country{Japan}}
\affiliation{%
  \institution{The University of Tokyo}
  \city{Tokyo}
  \country{Japan}
}
\email{itoh@comp.isct.ac.jp}

\author{Jun Rekimoto}
\orcid{0000-0002-3629-2514}
\affiliation{%
  \institution{The University of Tokyo}
  \city{Tokyo}
  \country{Japan}}
\affiliation{%
  \institution{Sony CSL Kyoto}
  \city{Kyoto}
  \country{Japan}}
\email{rekimoto@acm.org}

\renewcommand{\shortauthors}{Zhang et al.}

\begin{abstract}
    Presbyopia, a common age-related vision condition affecting most people as they age, often remains inadequately understood by those unaffected. To help bridge the gap between abstract accessibility knowledge and a more grounded appreciation of perceptual challenges, this study presents OpticalAging, an optical see-through simulation approach. Unlike VR-based methods, OpticalAging uses dynamically controlled tunable lenses to simulate the first-person visual perspective of presbyopia's distance-dependent blur during real-world interaction, aiming to enhance awareness. While acknowledging critiques regarding simulation's limitations in fully capturing lived experience, we position this tool as a complement to user-centered methods. Our user study (N = 19, 18-35 years old) provides validation: quantitative measurements show statistically significant changes in near points across three age modes (40s, 50s, 60s), while qualitative results suggest increases in self-reported understanding and awareness of perceptual challenges among participants. The integration of our tool into a design task showcases its potential applicability within age-inclusive design workflows when used critically alongside direct user engagement.
\end{abstract}



\begin{teaserfigure}
  \includegraphics[width=\textwidth]{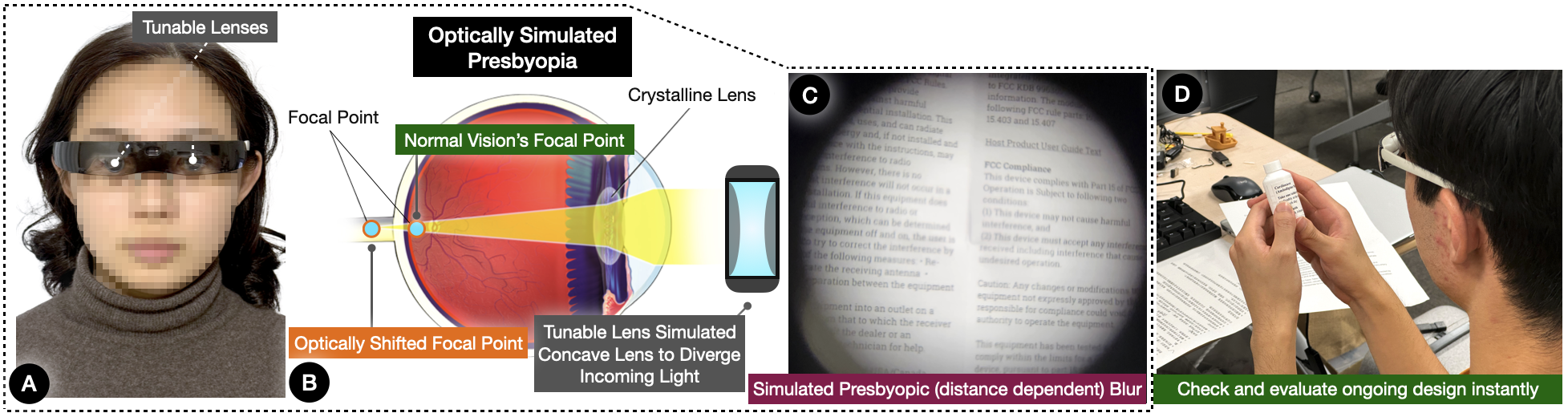}
  \caption{Optical simulation of presbyopia using tunable lenses. (A) A participant wearing the tunable lens eyewear. (B) Schematic diagram illustrating how the tunable lenses simulate presbyopia by optically shifting the focal point behind the retina, creating a presbyopic blur effect. (C) An example of simulated presbyopic blur is when viewing text at close range. (D) A designer using the simulation eyewear to evaluate and refine their design in real-time, demonstrating the practical application of the tool in the design process.}
  \label{fig:teaser}
\end{teaserfigure}

\maketitle

\section{Introduction}
Presbyopia, an age-related visual impairment, is a ubiquitous condition affecting nearly the entire aging population \cite{BMC2021Presbyopia, polat2012training, Charman2017Virtual, Wolffsohn2020Social, Heys2007Presbyopia, Atchison2008New}. Although traditional corrections such as spectacles or contact lenses offer functional solutions, they can introduce their own challenges \cite{katz2021presbyopia, Wolffsohn2023New}. For instance, constantly needing to switch between different pairs of glasses can be cumbersome, multifocal lenses often require an adaptation period and may introduce peripheral distortion, and contact lenses might be uncomfortable or impractical for certain activities or environments \cite{waghmare2022review}. This impact underscores a critical gap in inclusive design, which often stems from a lack of deep understanding among those unaffected by presbyopia \cite{sprung2022misunderstanding, singer2014understanding}. Addressing this understanding gap necessitates approaches that can help non-presbyopic individuals, particularly designers, appreciate the perceptual challenges associated with presbyopia, fostering broader awareness \cite{Goertz2014Presbyopia}. Bridging this gap requires tools that move beyond abstract design guidelines, helping designers develop a more grounded perspective than abstract knowledge alone might provide.

Advances in visual simulation technologies, including virtual reality (VR) and augmented reality (AR), have opened new avenues for simulating visual impairments. VR has shown versatility in simulating conditions and potentially enhancing understanding \cite{VRHealth2021, alexander2020design, vayrynen2016head}. However, while VR offers versatile simulation platforms, its inherent disconnect from the physical world limits its utility for crucial design evaluation tasks that involve interacting with physical prototypes, assessing ergonomics directly, or judging legibility on actual product surfaces and packaging within real-world environments. Optical see-through approaches, in contrast, allow simulation directly within the user's real-world context, enabling designers to experience simulated impairments while handling tangible mock-ups and interacting naturally with the intended use environment \cite{zhang2022seeing, krosl2020cataract, padmanaban2018autofocals, langlotz2018chromaglasses}.

Our research addresses this need by introducing OpticalAging, an optical see-through system using tunable lenses to dynamically simulate the visual effects of presbyopia during real-world tasks. This method aims to provide designers, welfare workers, and other related practitioners with a simulated, first-person perspective of presbyopia, intended to enhance understanding of specific visual challenges. Crucially, we recognize the limitations and critiques associated with using simulation for understanding disability \cite{bennett2019promise,nario2017crip}; OpticalAging is therefore presented not as a means to fully replicate lived experience, but as a tool to potentially complement direct engagement and collaborative design practices, potentially reshaping inclusive design for aging populations.

To validate our approach, we conducted a user study with three objectives: (1) quantitatively examine simulated presbyopia in three age modes (40s, 50s, and 60s); (2) assess reported changes in understanding and perspective-taking following interaction with the simulation; and (3) investigate the practical application of our simulation in real-world design scenarios with a professional designer and a design student. The results demonstrate that our approach significantly alters the near points of the users, enabling them to experience specific visual effects associated with presbyopia. Our system provides a perceptually relevant simulation of distance-dependent blur. Participants reported observations about daily challenges associated with these visual effects, suggesting increases in reported awareness and understanding. An exploratory case study with designer participants revealed the potential of our system to inform aspects of design processes, contributing towards more intuitive, user-centered solutions for aging populations.

This research contributes to the field of inclusive design support in several ways: (1) A novel optical see-through system using tunable lenses for first-person, real-time simulation of specific visual effects associated with presbyopia, enabling direct evaluation of real-world interactions. (2) A method intended to help bridge the gap between abstract accessibility guidelines and a more grounded appreciation of perceptual challenges relevant to inclusive design for aging populations. (3) Strong quantitative validation of the simulation's ability to accurately shift near points, complemented by qualitative evidence suggesting potential enhancements in reported understanding and perspective-taking among non-presbyopic users when interpreted within the limitations of simulation. (4) Explores the practical applications of presbyopia simulation in real-world design scenarios, offering insight into how such tools might be integrated into professional design workflows when used critically and supplementally.

\section{Related Works}

\subsection{Presbyopia: Background and Treatments}

    Presbyopia is primarily caused by decreased flexibility of the crystalline lens inside the eye as people age. Presbyopia typically manifests in individuals over 40 years old and affects a significant portion of the global population~\cite{Goertz2014Presbyopia}. In terms of self-awareness of presbyopia, patients with presbyopia often become aware of it in their late forties, but some may have difficulty with near vision-related tasks before becoming aware of it \cite{Tsuneyoshi2021Determination}. As the world population continues to age, we will see the rising prevalence of presbyopia turning into an urgent global issue.
    
    Current treatments for presbyopia, including reading glasses, bifocals, multifocal contact lenses, and surgical interventions, offer functional solutions but often fall short in addressing the complex visual challenges faced by presbyopia \cite{Grzybowski2020A, mayoPresbyopia}. Although these treatments provide vision correction, they come with limitations such as a reduced field of view, adaptation periods, and potential compromises in depth perception \cite{katz2021presbyopia}. Additionally, the trial-and-error approach often employed in prescribing these solutions suggests a gap in our understanding of the patient's visual experience \cite{moshirfar2022review}.
 
\subsection{Visual Impairment Simulation: From Static Filters to Optical See-Through}

Simulation techniques range from low-fidelity static filters \cite{ZEISS, asada, inclusivedesigntoolkitCambridgeSimulation} and full-body aging suits to Virtual Reality (VR) environments~\cite{jones2020seeing, krosl2020cataract}. While aging suits like the Third Age Suit~\cite{hitchcock2001third} and AGNES~\cite{lavalliere2017walking} effectively simulate reduced joint mobility and tactile loss, their visual components typically rely on static spectacles (e.g., yellow tints or fixed blur). Conversely, VR systems can render complex aberrations like glare and field loss~\cite{alexander2020design, orlosky2024eye}, but they rely on video pass-through (VST). This introduces latency and creates a disconnect from the physical world, limiting their utility for evaluating tangible prototypes where haptic feedback and focal depth are critical.

To address these limitations, researchers have turned to Optical See-Through (OST) approaches. For instance, Zhang et al.~\cite{zhang2022seeing} utilized OST eyewear to simulate glaucoma, emphasizing the value of an unmediated view of reality. However, prior OST works and static aging suits have not adequately addressed the dynamic, distance-dependent blur characteristic of presbyopia. Capturing this requires a real-time, mechanically adjustable optical system that goes beyond static diffusion.

Our approach specifically targets this gap to enable high-fidelity physical interaction studies. Unlike VST systems which may induce motion sickness or obscure fine surface details (e.g., paper texture), OpticalAging allows designers to assess legibility and ergonomics in their intended environment. Crucially, we situate this tool within the ethical framework proposed by disability scholars \cite{tigwell2021nuanced, bennett2019promise}: we present simulation not as a replacement for lived expertise, but as a pragmatic ``bridge'' for designers operating under constraints of time and access, designed to highlight physiological barriers rather than replicate the full psychosocial experience of disability.

\subsection{From Corrective Devices to Simulation Tools: A New Application for Tunable Lenses}

Foundational work by researchers like Hasan et al. \cite{hasan2017tunable}, Li et al. \cite{li2006switchable}, and others \cite{chen2023electrically, he2019recent, liu2023tunable, milton2014electronic, Jiang2021Electrohydrodynamic} has detailed the optical principles and control algorithms for these lenses. Building on this, applied research from Mompeán et al. \cite{mompean2020portable} and Padmanabhan et al. \cite{padmanaban2018autofocals} demonstrated the efficacy of these lenses for dynamic presbyopia correction. Most recently, Agarwala et al. \cite{agarwala2022evaluation} evaluated liquid membrane-based systems for aberration control and feedback mechanisms. Zhang et al. \cite{zhang2024aged} demonstrated the psychophysical feasibility of using the ViXion01 eyewear to optically shift near points. Our work adopts this hardware framework but moves beyond technical feasibility to evaluate the system's ecological validity and its utility as a design probe—areas not addressed in previous quantitative studies.

While Zhang et al. established technical feasibility, their work was limited to  optical measurements in controlled settings~\cite{zhang2024aged}. The ecological validity of such systems, whether they produce perceptually meaningful effects during real-world tasks—and their practical utility as design probes remain unexplored. Our work addresses this gap by providing mixed-methods evaluation and an exploratory design deployment.

\section{Simulating Presbyopia through Dynamic Adjustment of Diopters using Tunable Lenses}

\begin{figure}
    \centering
    \includegraphics[width=\linewidth]{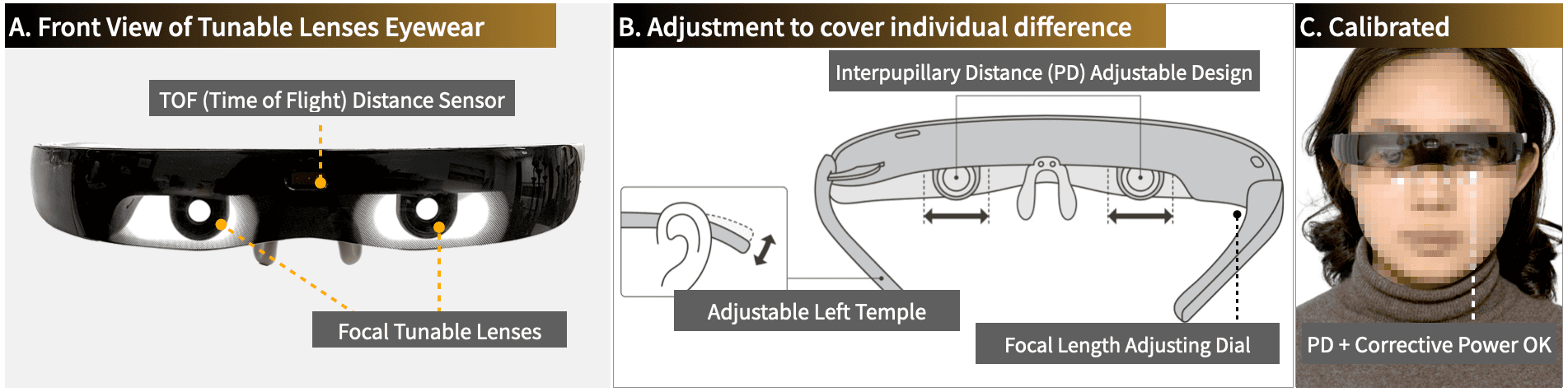}
    \caption{Configuration and calibration of the tunable lenses eyewear for presbyopia simulation. (A) Front view of the eyewear, showing the placement of the focal tunable lenses and the Time of Flight (TOF) distance sensor. (B) Adjustable features of the tunable lens eyewear to accommodate individual differences, including interpupillary distance (PD) adjustment, adjustable left temple, and a focal length adjusting dial. (C) A participant wearing the calibrated eyewear, with proper PD and corrective power settings applied.}
    \label{fig:hardware}
\end{figure}

\begin{table}[h]
    \centering
    \small 
    \caption{Tunable Lenses (ViXion01) Specifications.}
    \label{tab:lens_specs}
    \begin{tabularx}{\columnwidth}{>{\centering\arraybackslash}X >{\centering\arraybackslash}X >{\centering\arraybackslash}X >{\centering\arraybackslash}X >{\centering\arraybackslash}X}
        \toprule
        \textbf{Dimensions (Ø$\times$Thick)} & \textbf{Optical Power} & \textbf{Clear Aperture} & \textbf{Op. Temp.} & \textbf{Pupillary Dist.} \\
        \midrule
        15.8 $\times$ 7.4mm & -15D $\sim$ 15D & 6.3mm & 0 $\sim$ 45$^{\circ}$C & 55 $\sim$ 70mm \\
        \bottomrule
    \end{tabularx}
\end{table}

    Our approach optically simulates presbyopia by dynamically adjusting two tunable lenses to shift the near point of the user to an intended aged condition (e.g., 40s, 50s, and 60s modes in this research). The near point here is the one that represents the amplitude of accommodation instead of the near point of convergence. The tunable lenses are built-in ViXion01\footnote{\url{https://vixion.jp/en/vixion01/}}, a pair of commercially available autofocus smart eyewear, its specifications are listed in Table \ref{tab:lens_specs}. It should be noted that this specific hardware platform, like many wearable displays, has a limited field of view, a factor discussed further in the Limitations section. This eyewear was initially designed to work as corrective glasses, helping its wearer see both closed and distant objects clearly. Our approach, on the other hand, uses the tunable lenses of the eyewear to optically simulate two of the primary visual representations of presbyopia: (1) Increased near point of accommodation along with aging, which appears in the tendency of older adults to hold the reading material farther away to make the visual content clearer\footnote{\url{https://www.mayoclinic.org/diseases-conditions/presbyopia/symptoms-causes/syc-20363328}}\cite{rozanova2018fundamentals,truscott2009presbyopia,atchison1995accommodation}. (2) Blurred vision within the near point range~\cite{patel2007presbyopia}.

\begin{figure*}
    \centering
    \includegraphics[width=0.9\linewidth]{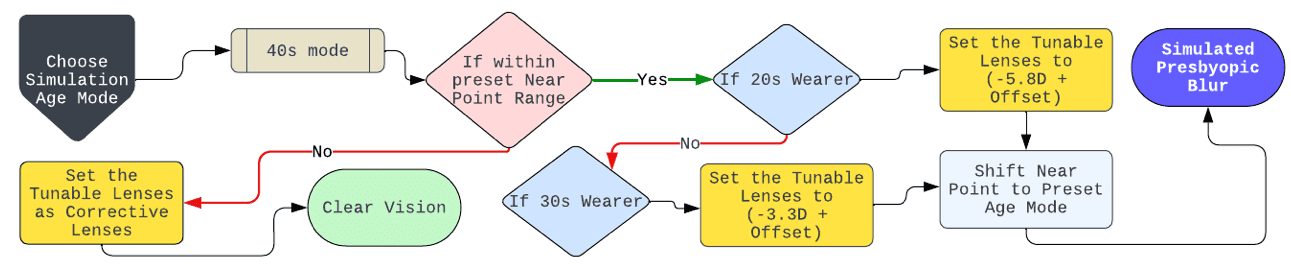}
    \caption{Workflow of the presbyopia simulation algorithm. The flowchart illustrates the decision-making process for simulating presbyopic blur based on the chosen age mode, wearer's age, and viewing distance relative to the preset near point range.}
    \label{fig:algorithmFlowChart}
\end{figure*}

    \textbf{Algorithm: }As presbyopia gradually progresses with age and individuals typically begin to notice symptoms in their 40s \cite{duane1922studies, NIH2016}, our simulation modes commence at this age, extending through the 50s and 60s, in line with previous research \cite{duane1912normal, schwartz2013geometrical}. To simulate presbyopic visual blur, the adjustable lenses alter the cumulative focal power of the wearer. The algorithm operates simply: If the tunable lens eyewear's distance sensor identifies that the viewing plane is nearer than the preset near point (e.g., in the 40s, 50s, or 60s mode), the lenses adjust their diopters to decrease the wearer’s overall accommodation ability. This logic is the inverse of a corrective autofocus system. Whereas a corrective system would add positive diopter power to aid near-vision focus, our algorithm intentionally applies a negative diopter offset (e.g., -5.8D for a 20s user in 40s mode) precisely when the user attempts to focus up close. This computationally shifts the focal point behind the retina, simulating the physiological effect of presbyopia rather than correcting it. This results in the intended presbyopic blur. The detailed workflow can be found in Figure \ref{fig:algorithmFlowChart}.

\begin{figure*}
    \centering
    \includegraphics[width=\textwidth]{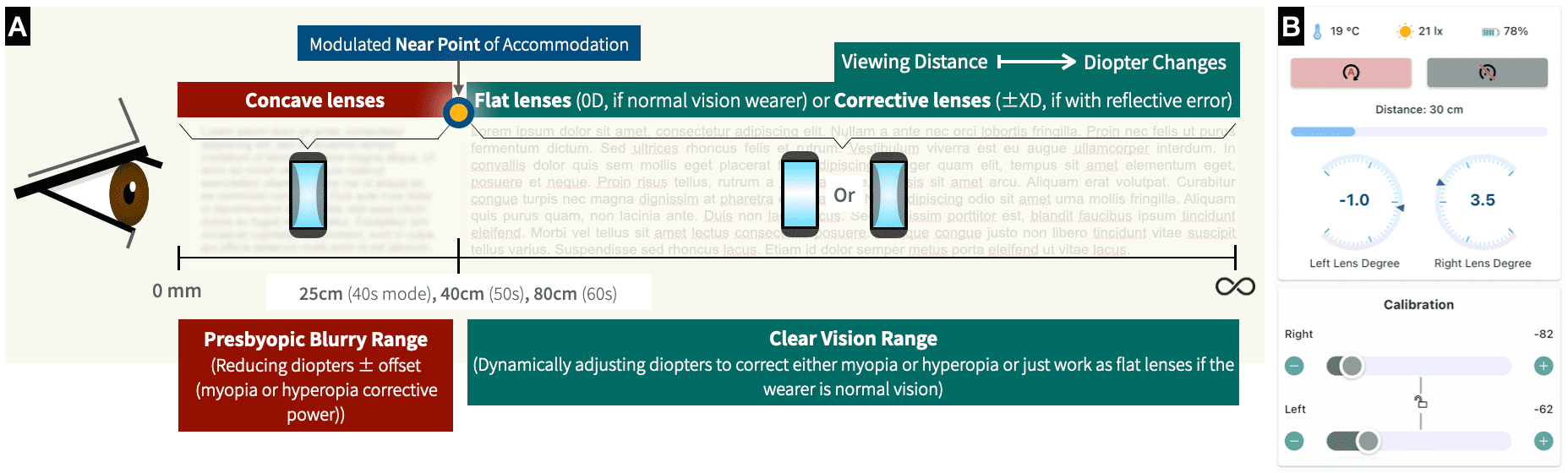}
    \caption{Approach and interface for presbyopia simulation using tunable lenses. (\textbf{A}) Illustration of the primary algorithm: The tunable lenses simulate presbyopic blur by functioning as concave lenses to reduce the overall diopter of the wearer when the viewing distance is less than a preset near point (e.g., 25 cm for 40s mode). Beyond this point, the lenses work as corrective mode to adjust dynamically based on viewing distance. (\textbf{B}) Operational graphical interface for individual calibration, allowing separate focal power correction for each eye to accommodate wearers with refractive errors.}
    \label{fig:logicFlow}
\end{figure*}

    The choice to simulate presbyopia across the age groups of the 40s, 50s, and 60s is based on physiological changes and clinical importance. Typically, the 40s signify the beginning of noticeable presbyopic symptoms, the 50s denote a crucial midpoint in the condition's progression, and the 60s generally show peak presbyopia \cite{duane1912normal, duane1922studies, NIH2016}. This age range highlights essential phases in the deterioration of lens elasticity and ciliary muscle effectiveness, the main causes of presbyopia \cite{KPierscionek1993What, Atchison2008New}. Additionally, each decade signifies different stages of how presbyopia affects daily tasks and workplace efficiency \cite{Kanasi2016The}.

    To initialize the presbyopic simulation, tunable lenses are precisely individually adjusted to operate as concave lenses, adjusting the near point of accommodation of the wearer. This adjustment alters the overall diopter strength of the user's vision to a preset simulation mode. The specific diopter values used in our simulation modes (Table \ref{tab:ageTodiopterValues}) are derived from the seminal accommodation amplitude data from Duane \cite{duane1912normal, duane1922studies, purves2008neuroscience}. Our algorithm operationalizes this clinical data into a dynamic, interactive system by calculating the required negative diopter adjustment needed to reduce a user's baseline accommodation to match the target age's average near point.

    For instance, in the case of a 20-year-old wearer (assuming an amplitude of accommodation (AoA) of 9.75D) with normal vision in the 40s simulation mode, the tunable lenses act as 0D lenses when the focused plane is farther than 25cm, which enables the wearer to see things clearly as usual. When the distance from the focus plane is shorter than 25cm, the tunable lenses function as -5.75D concave lenses. This consequently reduces the overall amplitude of accommodation to 4.0D (the average for people in their 40s) and shifts the focal point to a location beyond the wearer's retina.
    
    In contrast, in the case of simulating 40s, if the 20s wearer is nearsighted, the tunable lenses function as corrective lenses when the distance from the focused plane is greater than 25cm away from the wearer. Their corrective focal power here is stored as an \textit{offset}. The distance between the tunable lens eyewear and the object viewed determines how the diopter strengths of the tunable lenses are adjusted. The varied tunable lenses' states are illustrated in Figure~\ref{fig:logicFlow}. The distance from the viewing plane is continuously measured by a built-in Time-of-Flight sensor of the tunable lenses' eyewear. This distance value is converted to the diopter adjustment for each lens to support the wearer with clear vision in real-time. When the distance between the focused plane and the tunable lens eyewear is shorter than 25cm, the tunable lenses now function as a power of ``-5.8D + \textit{offset}'' to shift the focal point of the wearer outside the retina to simulate presbyopia. The corresponding diopter settings of other age modes and wearer age combinations, calculated using the same principle illustrated for the 20s and 30s examples in Figure \ref{fig:algorithmFlowChart}, are described in Table \ref{tab:ageTodiopterValues}.

\begin{table}
    \centering
    \caption{The diopter values employed to adjust the tunable lenses presented in this table are based on the research of Duane \cite{duane1912normal} and Schwartz \cite{schwartz2013geometrical}. These values are rounded to one decimal place to match the precision of the tunable lenses in our method.}
    \label{tab:ageTodiopterValues}
    \begin{tabularx}{\columnwidth}{c >{\centering\arraybackslash}X >{\centering\arraybackslash}X >{\centering\arraybackslash}X} 
        \toprule 
        \textbf{Wearer} & \multicolumn{3}{c}{\textbf{Target Presbyopic Mode}} \\
        \cmidrule(lr){2-4} 
        \textbf{Age} & \textbf{40s} & \textbf{50s} & \textbf{60s} \\ 
        \midrule 
        20s & -5.8D & -7.3D & -8.5D \\ 
        30s & -3.3D & -5.1D & -6.0D \\ 
        \bottomrule
    \end{tabularx}
\end{table}

\textbf{Accommodation Speed Control: }Our approach implements a non-linear accommodation speed control for the tunable lenses to mimic the accommodation speed and process observed in the aging human eye \cite{Schaeffel1993Inter‐individual}. The algorithm uses a non-linear interpolation between current and target focal lengths. Key to this process are two variables: a time constant that adjusts the speed of accommodation, and a delta time based on the system's refresh rate.

The interpolation uses an exponential decay function, allowing for smooth, gradual adjustments that are less perceptible and potentially more comfortable for the user.

\textbf{Debounce: }To ensure stable distance measurements and avoid sudden changes in lens adjustments, we incorporated a debounce algorithm. This algorithm maintains a circular buffer that holds the latest distance readings, with configurable options for the number of samples and the consensus threshold. For every new distance reading: (1) Insert the new value into the circular buffer while removing the oldest value, (2) calculate the frequency of each distinct distance value in the buffer, (3) if any distance value reaches or surpasses the predefined threshold, it is deemed the stable reading, (4) if no value meets the threshold, the new distance is accepted as is.

    \textbf{Temperature Compensation Algorithm for Focal-Tunable Lenses:}
    Considering that temperature changes can impact the performance of focal-tunable lenses due to the thermal sensitivity of the liquid within the lens, our algorithm dynamically adjusts the lens settings in response to temperature shifts to maintain consistent focal properties under varying conditions.

\subsection{Replication and Generalizability}

    While the specific hardware platform used in our study is subject to a Non-Disclosure Agreement (NDA), our methodology and algorithmic principles are generalizable and designed for replication. The core components—commercially available tunable lenses and a distance sensor—can be integrated following the logic detailed in this paper.

    The critical element for replication is the implementation of an algorithm that, based on distance input, applies a negative diopter adjustment to reduce the user's accommodative amplitude to a desired, age-specific level (see Table \ref{tab:ageTodiopterValues}), rather than a positive one for correction. To further aid replication, we have provided a detailed calibration and simulation workflow in our flowcharts (Appendix Figure \ref{fig:calibrationFlowChart} and Figure \ref{fig:algorithmFlowChart}). We will also make our implementation notes and detailed algorithmic logic (e.g., pseudo-code) available as supplemental material to ensure the research community can build upon our work.

\section{User Study}
    To explore and validate our approach, our user study was structured into three distinct parts. Part I involved the evaluation of optically shifted near points for simulated presbyopic age groups corresponding to the 40s, 50s, and 60s. Part II required participants to perform 20-minute daily reading activities with our simulation eyewear in the experimental environment. Finally, Part III was an exploratory N=2 case study with a professional product designer and a design student. The goal of this final part was not to produce generalizable findings, but to generate initial, qualitative insights into the practical application of our simulation and to illustrate its potential use within a professional design workflow. Our research protocols were rigorously reviewed and have obtained approval from the Institutional Review Board (IRB) of our university.
    
    It is important to note that the design of these main study components was informed by prior exploratory work. Specifically, before conducting these main studies, we conducted preliminary qualitative explorations with a small group of older adults (N = 4, ages 50s-60s) experiencing presbyopia. The goal was to gather initial insights into their subjective visual experiences to better inform the development and evaluation of the OpticalAging simulation. These informal discussions provided valuable early feedback on the nature of presbyopic blur and near point variability.

    \textbf{Calibration and Individual Adjustment.}
    The user study began with a careful calibration procedure to tailor to each participant's unique visual needs. Using tunable lenses' ability to simulate concave or convex lenses, nearsighted or farsighted participants participated in the study without their regular corrective eyewear. The corrective values determined by the participants were then incorporated into our simulation algorithm as the \textit{offset}.
    
    First, the participant puts on the ViXion01 eyewear and adjusts the interpupillary distance to position the tunable lens modules precisely in front of their eyes; see Figure \ref{fig:hardware} for a detailed hardware structure. Following ViXion01's official guidelines, and our calibration workflow, see Figure \ref{fig:calibrationFlowChart}, we ensure that the embedded tunable lenses initially functioned as corrective lenses to counteract potential nearsightedness or farsightedness. If the wearer has normal vision, the tunable lenses function as 0D lenses so that they can see things clearly as usual, when the focal plane is far away from the preset near point range. Participants who reported having no noticeable visual acuity difference between the two eyes are instructed to manipulate the adjustable dial, which changes the lenses' diopters until clarity is achieved for a standing hardcover book at a distance of one meter. Individual lens adjustment is also performed for participants reporting anisometropia (disparate visual acuity between their eyes), Figure \ref{fig:calibrationFlowChart} demonstrates the detailed calibration workflow. In this individual adjustment case, all calibration steps were remotely monitored and controlled using ViXion01's official smartphone application, the graphical user interface is illustrated in Figure~\ref{fig:logicFlow} B. 
    
    \begin{figure*}[t]
    \centering
    \includegraphics[width=0.8\textwidth]{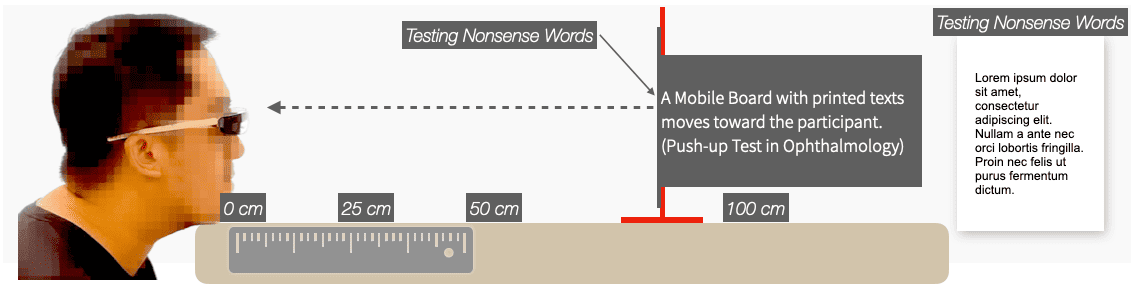}
    \caption{A participant wearing the tunable lens eyewear is shown with a mobile board displaying printed text moving towards them. This setup mimics the Push-Up test commonly used in ophthalmology to measure near point accommodation. The participant's jaw rests on a table to minimize head displacement, ensuring consistent and reliable measurements.}
    \label{fig:partIexperimentalEnvironment}
\end{figure*}

\subsection{Part I: Evaluation of Optically Shifted Near Points}

    This part aims to evaluate optically shifted near points for simulated age groups corresponding to the 40s, 50s, and 60s while using our optical-see-through approach. Although in ophthalmological tests, near point accommodation tests typically contain monocular and binocular conditions, we choose to examine binocular near point. Since most real-world near-vision tasks relevant to design evaluation (e.g., reading labels, using handheld devices, interacting with physical objects) are performed using both eyes, assessing the binocular near point provides a more ecologically valid measure for simulating the functional impact of presbyopia in everyday scenarios.
    
    \textbf{Participants:} We recruited 19 individuals from diverse professional backgrounds for this study via word of mouth. They ranged from 18 (with consent from the individual and their guardian) to 35 years old, with an average age (MEAN) of 27.8 years and a standard deviation (SD) of 4.1. We specifically targeted individuals under 40 years old, as those in their 40s typically begin to experience presbyopia \cite{duane1922studies}. All participants were provided with essential information about this user study through a consent form. To ensure participant safety, we adopted exclusion criteria consistent with prior aging simulation studies~\cite{lavalliere2017walking}. Participants were screened for history of dizziness, balance disorders, or recent falls. Additionally, the maximum continuous wear time was limited to 20 minutes to prevent disorientation or motion sickness. They took part in the study after signing the consent form. As for the baseline eyesight criteria, because the ViXion01's built-in tunable lenses can optically correct both nearsightedness and farsightedness, participants were allowed to participate without wearing corrective glasses or contact lenses. Detailed calibration regarding the fitting of their interpupillary distance and corrective focal power is described in the previous \textit{Approach} Section. Each participant received a 1000 JPY Amazon gift card as compensation for their time.
\begin{figure*}
    \centering
    \includegraphics[width=0.95\textwidth]{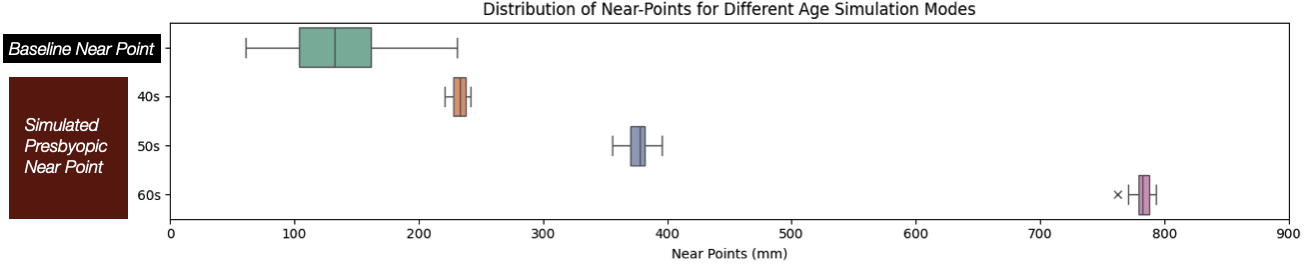}
    \caption{This plot includes data for the baseline (no simulation, natural near point) and simulated conditions representing optically shifted near point of 40s, 50s, and 60s for 19 participants. Each box represents the interquartile range (IQR) with the median indicated by the central line. Whiskers extend to the minimum and maximum values within 1.5 times the IQR. Outliers, marked with `x', are data points beyond the whiskers. The plot demonstrates a clear progression in near point distances as the simulated age increases. Statistical analysis confirmed significant differences between all simulation modes ($p < .001$ for all pairwise comparisons).}

    \label{fig:horizontalPlot}
\end{figure*}

    \textbf{Procedures: }Following calibration, participants underwent (i) a baseline near point measurement, (ii) 40s, (iii) 50s, and (iv) 60s simulation modes in total, with roughly one to two minute breaks in between each pair of conditions. The order of these four conditions was counterbalanced. The experimental setup was situated in an art studio with artificial lighting. Participants undertook the task one after another with a five-minute interval between each. As illustrated in Figure \ref{fig:partIexperimentalEnvironment}, participants positioned their jaw on a table to limit unnecessary head movements. A board with English letters printed in a Lorem ipsum format (nonsense text) moved toward the participant and paused when the letters became consistently blurry, as reported by the participants. This method draws on the widely used Push-Up and Pull-Away techniques \cite{hamasaki1956amplitude, esmail2016comparison} in the field of ophthalmology. To avoid false positives due to blur induced by eye strain, both the current diopter settings of the tunable lenses and the distance measured by the tunable lens eyewear's built-in TOF sensor were checked for accuracy.

    \textbf{Baseline Near Point Measurement: }In establishing a baseline for evaluating our approach to simulating various shifted near points, we use our tunable lenses eyewear in corrective glasses mode or in flat lenses mode (if normal vision) for this baseline measurement. This decision was based on two key factors: (1) Ensure that all visual assessments are made through the same optical system. (2) Allows us to measure the ``best corrected'' near point for each participant, providing a more standardized baseline across our study population.

\subsection{Part I Findings: Validation of Optically Shifted Near Points}

The quantitative evaluation confirms that OpticalAging effectively and significantly modulates the user's near point of accommodation. As shown in Figure \ref{fig:horizontalPlot}, we observed a distinct, non-linear progression in near point distances corresponding to the simulated age increments. A repeated measures ANOVA (Greenhouse-Geisser corrected) revealed a significant main effect of simulation mode on near point measurements ($F(1.09, 19.64) = 2365.2, p < .001$) with an exceptionally high effect size ($\eta_p^2 = 0.992$). This magnitude of effect validates the system's capability to enforce a highly controlled physiological constraint, effectively overwriting the user's natural accommodation.

Post-hoc analyses (Bonferroni correction) confirmed significant differences between all condition pairs ($p < .001$). Specifically, the median near point shifted from a baseline of 132 mm to 233 mm (40s), 378 mm (50s), and 782 mm (60s). These measured values align closely with the theoretical presbyopia models defined by \cite{duane1922studies, NIH2016, schwartz2013geometrical} that informed our algorithm. Crucially, the low inter-participant variability—persisting across diverse baseline visual conditions (normal, myopia, post-LASIK)—demonstrates the system's robustness in delivering a standardized impairment experience. While minor deviations from algorithmic targets were observed, likely attributable to the system's debounce logic and measurement precision, the simulation successfully replicates the functional deficit required for design evaluation.

\subsection{Part II: Subjective Experience of Simulated Presbyopia (50s Mode) During Daily Reading Tasks}

    To complement our quantitative measurements of shifted near points, we designed this qualitative assessment to evaluate the subjective experience of participants using our presbyopia simulation system in daily tasks. \textbf{Participants} were in the same group as in Part I, they joined this part after a 5-minute break. Regarding these task-based assessments, specifically the reading tasks, the simulation was set to the 50s mode. This decision was based on several key factors: (1) Progression of Presbyopia: By the 50s, presbyopia is typically well-established in most individuals \cite{truscott2009presbyopia}. (2) Significant Impact on Reading: The 50s typically mark a point where presbyopia significantly affects near vision tasks, especially reading. 

    (3) Relevance to Target Users: Many users of presbyopia-simulating technology are likely to be younger individuals or designers seeking to understand the experience of more advanced presbyopia. 

    (4) Comparative Baseline: For younger participants, this mode provides a clear contrast to their natural vision, allowing for meaningful comparisons and insights into the challenges faced by individuals with presbyopia.
\begin{figure*}
    \centering
    \includegraphics[width=0.95\linewidth]{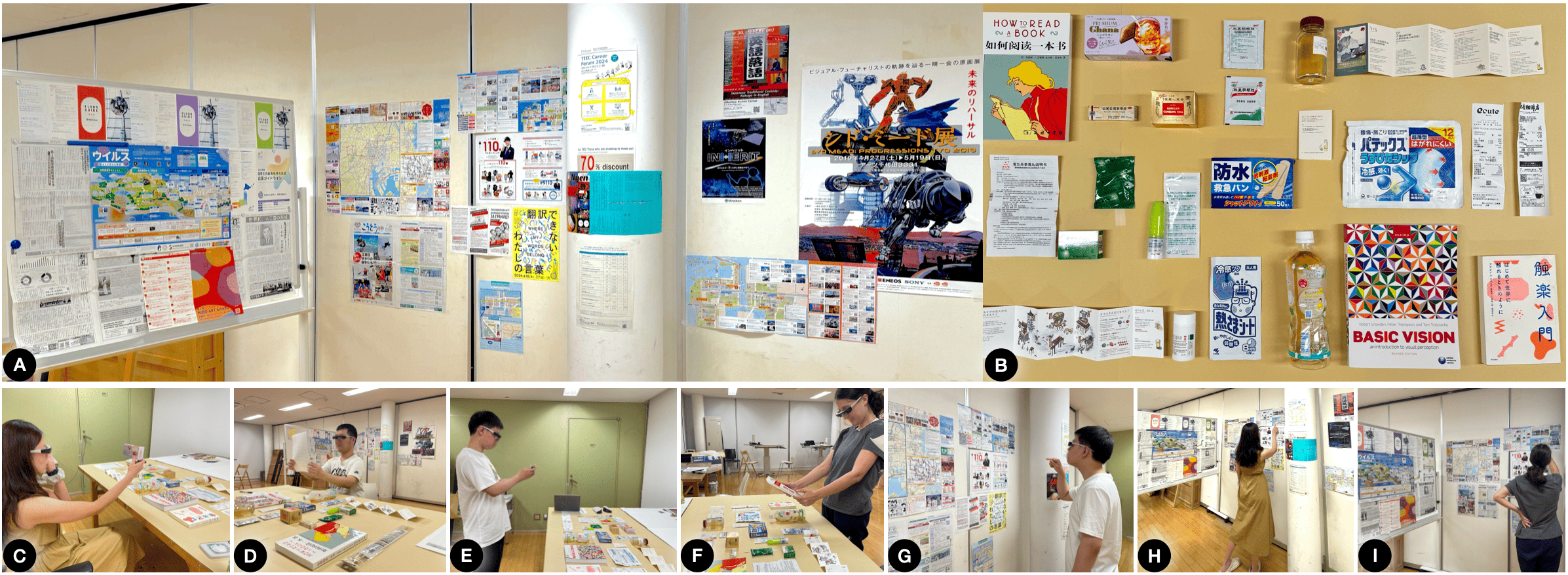}
    \caption{Experimental setup and participant engagement for Part II of the study. (A-B) Experimental environment and visual stimuli, including diverse reading materials such as posters, maps, product packaging, and instructional materials. (C-I) Actual experimental scenes showing participants interacting with various stimuli while wearing the presbyopia simulation eyewear. Participants can be seen examining different types of text and visual information at various distances and angles, simulating real-world scenarios encountered by individuals with presbyopia.}
    \label{fig:readingMaterials}
\end{figure*}
\begin{figure*}
                \centering
                \includegraphics[width=0.9\linewidth]{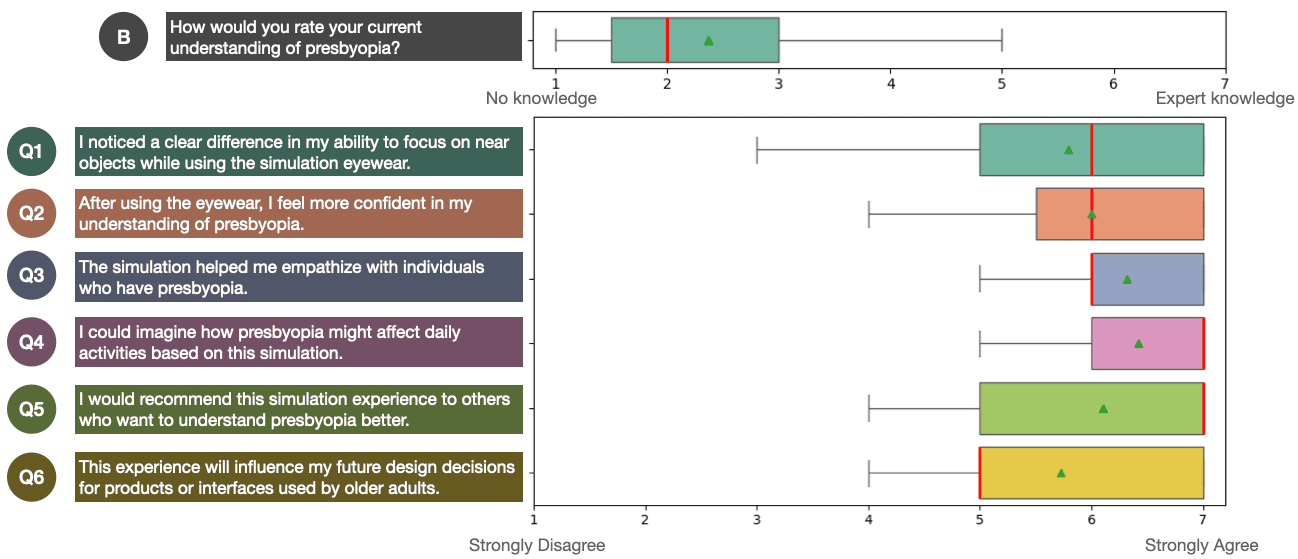}
                \caption{This figure presents boxplots showing the distribution of participant responses to a baseline question (\textbf{B}) and six statements (Q1-Q6) about their experience with the presbyopia simulation. The baseline question (\textbf{B}) assessed participants' initial understanding of presbyopia before the simulation. Q1-Q6 evaluated various aspects of the simulation experience after completing tasks with the 50s simulation mode. Responses were recorded on a 7-point Likert scale. Each box represents the interquartile range (IQR) with the median indicated by a vertical line. Whiskers extend to the minimum and maximum values within 1.5 times the IQR. Green triangles represent mean values. The red vertical lines indicate the median for each question. Key findings include: (B) low initial understanding of presbyopia (median = 2), (Q1) clear perceived difference in near vision focus (median = 6), (Q2) increased confidence in understanding presbyopia (median = 6), (Q3) enhanced perspective-taking regarding individuals with presbyopia (median= 6), (Q4) improved ability to imagine presbyopia's impact on daily activities (median = 7), (Q5) high likelihood of recommending the simulation (median = 7), and (Q6) potential influence on future design decisions for products used by older adults (median = 5, n = 11 participants with design backgrounds). The plot demonstrates a substantial shift from low initial understanding to high agreement across all aspects of the simulation experience.}
                \label{fig:partii-questionnaire-results}
            \end{figure*}
    
    \textbf{Procedures: }
    Initially, participants responded to a question gauging their baseline knowledge to evaluate their prior comprehension of presbyopia before the simulation. Afterward, they had 20 minutes to complete a range of near-vision tasks. These tasks involved reading printed texts, engaging in smartphone activities (unlocking, checking emails, replying to messages, web browsing, and other routine tasks), and walking while observing their environment. The reading materials were preselected in diverse languages to reflect the different cultural backgrounds of the participants. These included books, medicine bottles, medical instructions, food packaging, beverages, and posters within the experimental setup, as illustrated in Figure \ref{fig:readingMaterials}. The reading materials were chosen for their importance in conveying everyday information. For instance, the text on a medicine bottle or its specifications often details medical uses and possible side effects, which are crucial to heed. Similarly, food packaging information, such as allergen alerts and expiration dates, is vital in daily life. Furthermore, items like books, posters, receipts, and flyers are typical daily reading sources. These activities were specifically chosen for their high ecological validity, representing common daily tasks where presbyopia can significantly hinder interaction and access to information, thus providing a realistic context for evaluating the subjective experience of the simulation. An example of the reading materials is depicted in Figure \ref{fig:readingMaterials}. The experimenters utilized a checklist to confirm that all reading materials had been reviewed. To substantiate that participants explored and read the items, they were required to vocalize specific information printed on each item, such as typical medical uses and side effects from a medicine bottle or allergen warnings and expiration dates from food packaging.
    
    After finishing their tasks, participants answered the questions using the 7-point Likert scale questionnaire to evaluate their comprehension of presbyopia following the use of the optical simulation method. The questionnaire consists of seven structured questions (Figure \ref{fig:partii-questionnaire-results}) and two open-ended questions: (1) What aspect of the presbyopia simulation did you find the most surprising or insightful? (2) What would you suggest to further improve this presbyopia simulation approach? The questions are designed to gather additional feedback on their experiences with the simulation system, offering qualitative insights. The responses to the open-ended questions were documented in voice recordings with the participant's consent.

\subsection{Part II Findings: Perceived Understanding and Perspective Taking}

The simulation significantly bridged the gap between abstract knowledge and embodied appreciation of presbyopic challenges. Participants initially reported low baseline knowledge (median $= 2$). Post-simulation, a Wilcoxon signed-rank test confirmed a statistically significant increase in their confidence regarding understanding presbyopia ($Z=-3.84, p<.001, r=0.62$). This quantitative shift was strongly corroborated by high scores in self-reported perspective-taking (Q3, median $= 6$) and the ability to imagine daily life impacts (Q4, median $= 7$), with a strong positive correlation between these two factors ($r_s=0.72, p<.001$).

Qualitatively, the simulation induced a visceral realization of the physical labor associated with presbyopia—a nuance often lost in static guidelines. Participants did not just report ``blur''; they reported the physical fatigue of compensation. As one participant noted regarding the need to extend their reach:
\begin{quote}
    ``It's especially challenging because you need to hold things far away and keep your hands steady, which makes them sore... I feel sympathy for older individuals [because] my hands become sore after holding items away for a few minutes.''
\end{quote}
This physical discomfort translated into emotional frustration and anxiety about their own future aging, suggesting that the tool effectively facilitates \textit{perspective-taking} through physiological constraint. However, we interpret these ``empathy'' scores cautiously; while the simulation successfully highlights immediate perceptual barriers and physical compensatory costs, it functions as a sensitizing tool rather than a replacement for the lived experience of long-term adaptation \cite{bennett2019promise}.
    
    \subsection{Part III: An Exploratory Case Study Evaluating Impact on Design Practice}

    While the previous parts of our study provided quantitative validation and subjective ratings, this final part is an exploratory case study with two participants (N=2). We must emphasize that due to the small sample size, the goal here is not to draw generalizable conclusions. Rather, the purpose is to provide a rich, qualitative illustration of how the OpticalAging simulation could be integrated into a design process and to generate initial hypotheses about its potential impact that can inform future, larger-scale research.
\begin{figure}
    \centering
    \includegraphics[width=\linewidth]{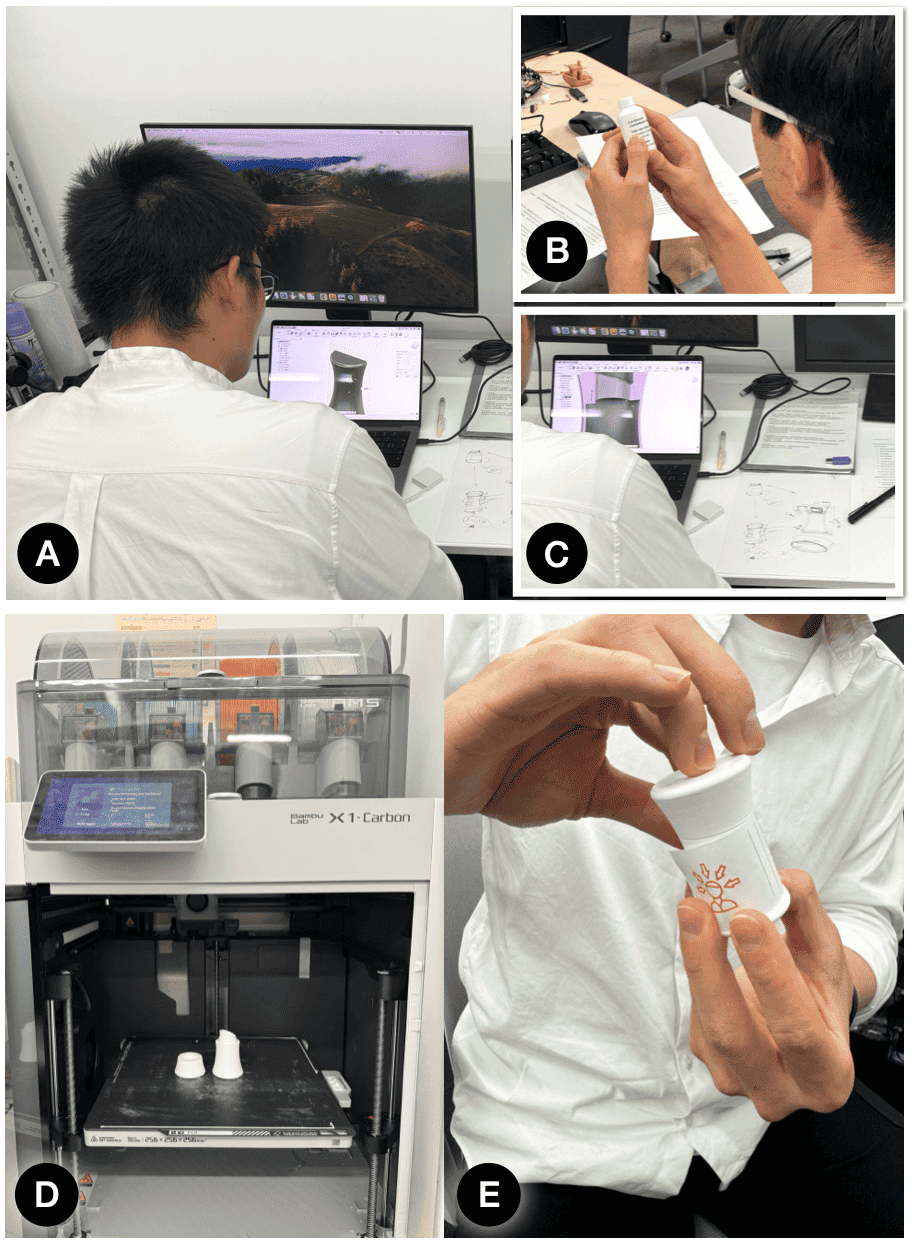}
    \caption{Exploratory case study experimental environment and design process. A) Professional designer creating the initial version of the medicine bottle design on a computer. B) Design student evaluating their current design plan using the presbyopia simulation eyewear. C) Professional designer performing section analysis of the bottle design. D) Rapid prototyping 3D printer used to materialize participants' design concepts, with the professional designer's first prototype freshly completed on the build plate. E) Professional designer examining the physical relationship between the 3D-printed bottle prototype and the printed label design.}
    \label{fig:out-of-lab-environment}
\end{figure}    

    We chose the design of a medicine bottle and its label as the central task. This task was selected because it represents a common real-world scenario where presbyopia can significantly hinder usability due to the need to handle the object, read fine print, and access critical health information (like dosage and warnings). It effectively combines challenges in both physical/ergonomic design and information design (layout, readability), making it a relevant and challenging task for exploring accessibility solutions.

    \textbf{Participants:} Two participants (male, 27 years old) were recruited, a professional product designer and a graduate design student. This pairing aimed to capture diverse perspectives, balancing industry experience with an academic viewpoint to understand the tool's potential applicability across different career stages and design contexts.

    \textbf{Simulation Mode:} The 50s simulation mode was used, maintaining consistency with Part II and reflecting a common age group for prescription medication users \cite{Kanasi2016The}, ensuring relevance to the design task \cite{Goertz2014Presbyopia, katz2021presbyopia}.

    \textbf{Procedure:} Participants first underwent the same calibration procedure as in previous tasks, using their personal laptops in a familiar environment (Figure \ref{fig:out-of-lab-environment}). We employed a within-participants design with two 45-minute conditions, separated by a 15-minute break. The first condition was always Free Design, where participants designed the presbyopia-friendly medicine bottle and label using their preferred methods, without our presbyopia simulation. In the second Simulation Condition, participants used the OpticalAging eyewear (set to 50s mode) to evaluate and redesign their initial concepts. In both phases, participants could use a provided 3D printer and laser printer to prototype their designs. We provided sample label content (medicine name, instructions). Designers considered bottle dimensions, capacity, ergonomics, label layout, typography, and color contrast. The primary goal given to participants was to design a bottle and label optimized for ease of use, particularly focusing on handling ergonomics and information legibility (dosage, warnings, etc.) for individuals likely experiencing presbyopia. While standard aesthetic considerations were relevant, functional accessibility for the target user group was the main priority. Following the tasks, participants completed a semi-structured interview and questionnaire assessing the simulation's ease of use, impact on design decisions, and overall usefulness.
    
    Although a direct comparison with other presbyopia simulation systems would be ideal, our extensive search revealed no currently available counterparts offering real-time, distance-dependent optical see-through simulation without requiring prior digitization of the environment (e.g., via 360$^{\circ}$ video).
   \begin{figure*}
        \centering
        \includegraphics[width=0.8\linewidth]{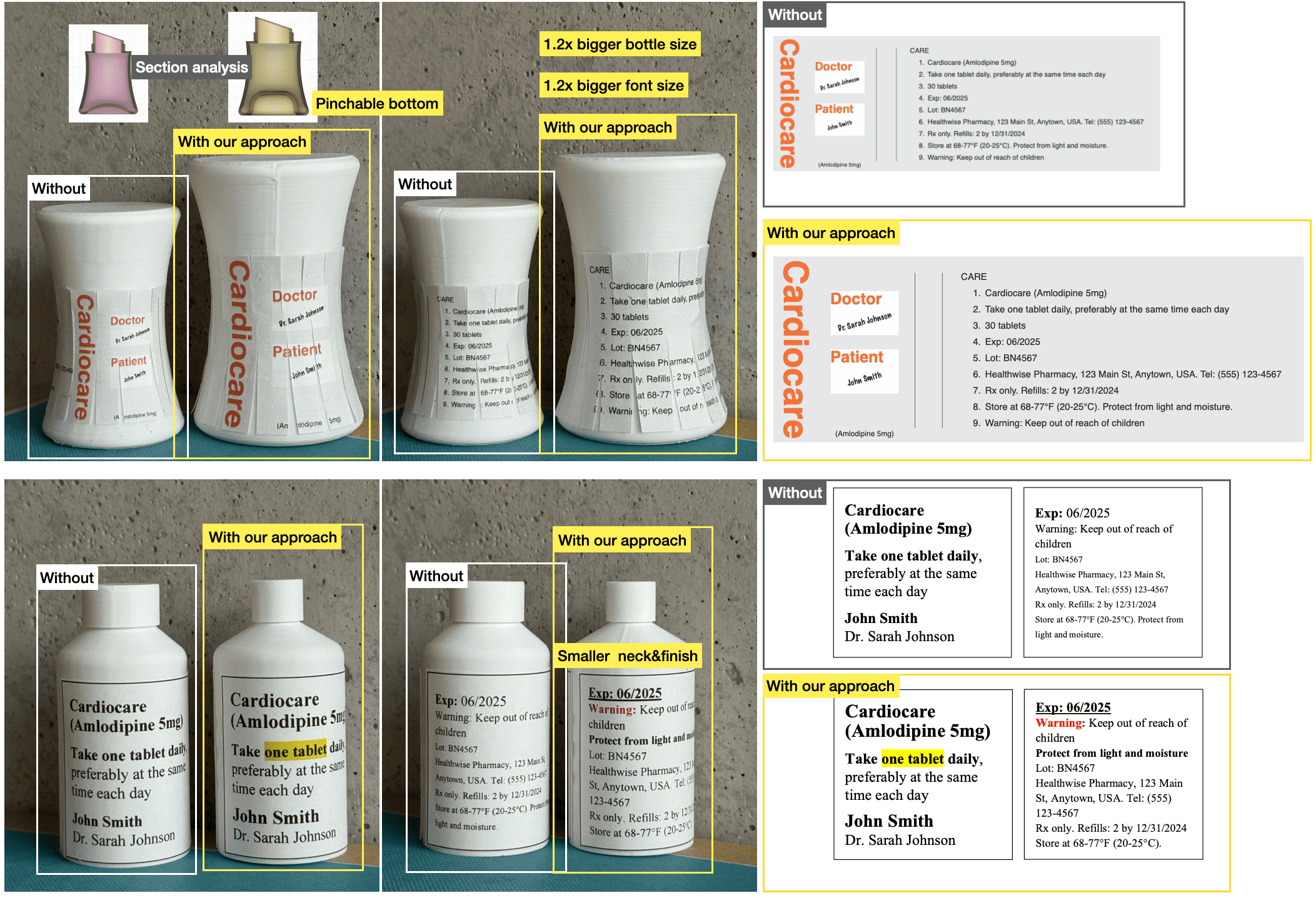}
        \caption{Comparison of medicine bottle designs with and without the presbyopia simulation approach. The top row shows the professional product designer's work, featuring a scaled-up bottle with a pinchable bottom and larger font sizes. The bottom row presents the design student's work, highlighting changes in bottle shape and label layout. Both sets of designs demonstrate modifications made after using the optical presbyopia simulation, including increased font sizes, adjusted layouts, and ergonomic improvements to enhance readability and usability for individuals with presbyopia.}
        \label{fig:productDesign}
    \end{figure*}

\subsection{Part III: Results}

        \textbf{Free Design Condition:} The designed medicine bottles and their labels are depicted in Figure \ref{fig:productDesign} within the white rectangles. For the professional designer, a trumpet-like opening was created to facilitate easier pill pouring for elderly users. Ergonomic considerations were integrated, with modifications to the bottle cap to ensure ease of opening, considering the potential decrease in hand strength in elderly users. For the design student's output, a generic bottle design was used as the base, with additional grooves added to the bottle body for better grip, though these were covered by the label. In the label design, the professional designer employed a multi-color scheme, with the medicine name (16pt), doctor and patient indicators (8pt) in orange, and other content in black (4pt) to emphasize key information with different visual hierarchy. Information was prioritized with the patient and doctor names given prominence for easy identification. The design student's label was monochromatic, with less noticeable font size differences among content on the label when compared to the professional designer. The medicine name was 11pt while the doctor and patient indicators were 10pt. 
        
        \textbf{Simulation Condition:} The professional designer's work included a 1.2x scaled-up bottle size, label size, and font sizes (16pt to 19pt). Additionally, there was an extra pinchable bottom design shown in the upper left of Figure \ref{fig:productDesign}. Similarly, the design student's work, shown in the bottom left of Figure \ref{fig:productDesign}, featured globally increased font sizes, with the medicine name increasing to 12.5pt in bold, the doctor name to 12pt in bold, and dosage and warnings highlighted with a yellow background and red respectively. While the bottle size did not increase under the optical presbyopia simulation for the student's design, as opposed to the professional designer, the size of the bottle neck and cap was reduced for improved dosage control.

\textbf{Questionnaire Results: }Both the professional designer and design student initially showed limited understanding of presbyopia (Baseline), each scoring 2 out of 7. Post-simulation feedback revealed high effectiveness across various design criteria. The approach was especially beneficial for evaluating and refining visual accessibility elements (Q4), with the professional designer giving it a perfect score of 7/7. Both participants found the method highly effective for selecting suitable typefaces (Q1), determining text sizes (Q2), and improving the overall design process (Q5), with all ratings ranging between 5 and 7. The simulation also proved valuable for choosing colors (Q3) and seamlessly integrating into existing workflows (Q6). Additionally, both participants expressed a strong inclination to use this approach in future designs (Q7), with the design student rating it 6/7 and the professional designer 5/7. These findings suggest that our simulation approach greatly improved the designers' capability to create accessible and user-friendly designs.

\textbf{Qualitative Results and Insights:}
    Emphasizing the professional designer's expertise, the feedback revealed that the most valued feature is the ability to instantly review the current design plan, which can greatly streamline the design iteration process. In the conventional design approach, needs of target users, such as individuals with presbyopia, were treated as a checklist, followed by a prolonged wait for client feedback on each iteration. Viewing from the actual presbyopic user's perspective enables agile iteration and improves the designer's confidence. When it comes to design requirements from clients or design handbooks, font size selection often follows vague guidelines like ``use a font size larger than or equal to a specified value.'' With simulation eyewear, he could refine the font size and style selection within a small range, balancing readability and density of information. The design student observed that, although larger texts had initially been selected, the simulation effectively underscored the necessity of further font size adjustments to better accommodate presbyopic patients. Both designers expressed that the first-person-view simulation enriched their understanding and perception of how everyday objects might appear to their parents and design audiences. This enhanced understanding led to a refined bottle design featuring a larger size, pinchable bottom, and reduced neck and cap diameter for improved dosage control. The design student also highlighted a difference between monocular and binocular viewing, noting that vision seemed much clearer (``as usual'') when looking through the simulation eyewear with only one eye open. This difference could be leveraged to facilitate a smooth transition from normal vision (single-eye viewing) to a presbyopic simulation (dual-eye viewing) without removing the simulation eyewear during design concept evaluations.
    
\subsection{Part III: Observations}

Given the limited sample size ($N=2$), we frame these findings strictly as preliminary observations intended to guide future inquiry rather than as generalized evidence. However, the distinct design modifications produced by both the professional and the student (Figure \ref{fig:productDesign}) point toward three potential utilities of the OpticalAging approach:

\begin{enumerate}
    \item \textbf{Facilitating Embodied Evaluation:} The simulation appeared to bridge the gap between abstract guidelines and immediate perception. The specific adjustments to font size and bottle ergonomics were driven by the designers' firsthand difficulty in reading and handling the prototypes. This aligns with Suri's concept of `designing in a state of empathy' \cite{suri}, suggesting that tunable lenses could serve as a mechanism for generating embodied insight during the drafting phase.
    
    \item \textbf{Potential for Rapid Iteration:} The participants' ability to instantly toggle between simulated presbyopia and corrected vision seemed to support a tighter feedback loop. This observation suggests that such tools might enhance process efficiency, allowing designers to ``spot check'' accessibility constraints in real-time without the logistical overhead of recruiting testers for every minor iteration.
    
    \item \textbf{Perceptual Nuances in Usage:} The feedback regarding monocular versus binocular viewing hints at opportunities for more sophisticated simulation interactions. Future work could explore how transitioning between these states might better mimic the fluctuating nature of visual fatigue.
\end{enumerate}

Ultimately, this case study illustrates that while simulation cannot replicate the lived experience, context, or compensatory strategies of aging, it may function as a valuable \textit{design probe}. It offers a ``first line of defense'' against inaccessible design, prompting immediate modifications (e.g., larger fonts, pinchable shapes) that might otherwise be overlooked until later testing stages. We emphasize that this tool is intended to complement, not replace, authentic user engagement with older adults.

\section{Discussion}

Our findings signal a paradigm shift in inclusive design methodology: moving from passive accessibility checklists to active embodied verification. Traditionally, accessibility guidelines function as external constraints---rules to be memorized and applied. However, as Hitchcock et al. argue, designers often disregard ergonomic data tables simply because they lack an intuitive grasp of the user's actual constraints~\cite{hitchcock2001third}. Our study supports this observation: designers using OpticalAging reported that the \textit{physical sensation} of visual fatigue---specifically the soreness in their arms from holding objects at a distance---prompted immediate design changes (e.g., pinchable bottles) that abstract guidelines had failed to inspire.

This shift in perspective is further evidenced by how participants attributed their difficulties. Similar to Lavalli\`{e}re et al.'s findings with the AGNES suit, where participants attributed difficulty to the ``store layout'' rather than the simulation tool itself~\cite{lavalliere2017walking}, our participants shifted their critique from their own vision to the \textit{design of the objects} (e.g., ``this font is too small'' rather than ``my eyes are blurry''). This indicates a successful shift toward inclusive design thinking, where the environment is viewed as the malleable variable rather than the user's ability.

Finally, while we acknowledge the critique that increased understanding may stem from information exposure rather than simulation alone, the somatic data collected suggests otherwise. The physical fatigue reported by participants serves as a unique data point that verbal briefings cannot convey. By injecting the physiological constraints of aging directly into the designer's visual stream, the system converts accessibility from a compliance task into an immediate, haptic reality.

    \subsection*{Limitations}
        
Several limitations should be considered when interpreting these findings: (1) \textbf{Simulation Fidelity:} As noted by critiques in disability studies~\cite{bennett2019promise, tigwell2021nuanced}, simulation risks oversimplification. Our system currently renders a uniform blur, which differs from the complex reality of aging vision that often includes `ghosting', scattering, or astigmatism~\cite{kleves2021prevalence, fresina2012myopia}. Therefore, our tool represents an idealized model of presbyopia (pure accommodative loss) rather than a comprehensive replication of the condition.

(2) \textbf{Study Scope and Generalizability:} The user studies focused solely on the 50s simulation mode and a short duration. Exploring onset (40s) or advanced (60s+) modes over longitudinal periods is necessary to understand long-term adaptation. Furthermore, the exploratory case study ($N=2$) serves only as an illustrative workflow example; larger-scale validation with diverse design tasks is required to generalize the impact on professional practice.

\section{Conclusion and Future Works}
    This paper presented OpticalAging, a real-time, optical see-through approach using tunable lenses to simulate the first-person visual perspective of presbyopia. Our findings, blending quantitative near-point validation with qualitative insights from designers, suggest that experiencing simulated presbyopic blur can significantly enhance understanding and awareness of perceptual challenges, potentially leading to more intuitive and user-centered designs for older adults. By providing a tool to help bridge the gap between abstract accessibility knowledge and embodied appreciation of perceptual challenges, this work offers a step towards more effective age-inclusive design practices.
    
    Looking ahead, the underlying tunable lens technology holds potential for simulating other visual conditions, such as cataracts or age-related macular degeneration. Future technical developments could include user-adjustable accommodation speed controls to further refine the simulation's utility in rapid design evaluations.
    
    By shifting the focus from correction to simulation, we demonstrate how these powerful devices can serve not only patients but also the designers, educators, and engineers working to build a more accessible world. Ultimately, this research contributes more than a novel optical hardware setup; it proposes a new interface for intergenerational understanding. By shifting the focus of tunable lenses from vision \textit{correction} to vision \textit{simulation}, we demonstrate how wearable displays can serve not just patients, but the architects of our daily environments. OpticalAging transforms the designer's own body into a high-fidelity proxy for the user, effectively reducing the cognitive distance between the `specification' of a disability and the `sensation' of living with it. While simulation can never fully replace the lived experience or compensatory strategies of older adults, providing designers with the ability to momentarily inhabit these perceptual constraints is a crucial step toward a world designed for the eyes of our future selves.

\begin{acks}
This work was supported by JST Moonshot R\&D Grant Number JPMJMS2012, JST ASPIRE Grant Number JPMJAP2327, and JST FOREST Grant Number JPMJFR206E; JSPS KAKENHI Grant Number JP20H05958 and JSPS Grant-in-Aid for Early-Career Scientists Grant Number JP25K21241; and JPNP23025 commissioned by the New Energy and Industrial Technology Development Organization (NEDO).
\end{acks}
\bibliographystyle{ACM-Reference-Format}
\bibliography{sample-base}


\appendix
\section{Appendix}

\begin{figure*}[t]
    \centering
    \includegraphics[width=0.9\linewidth]{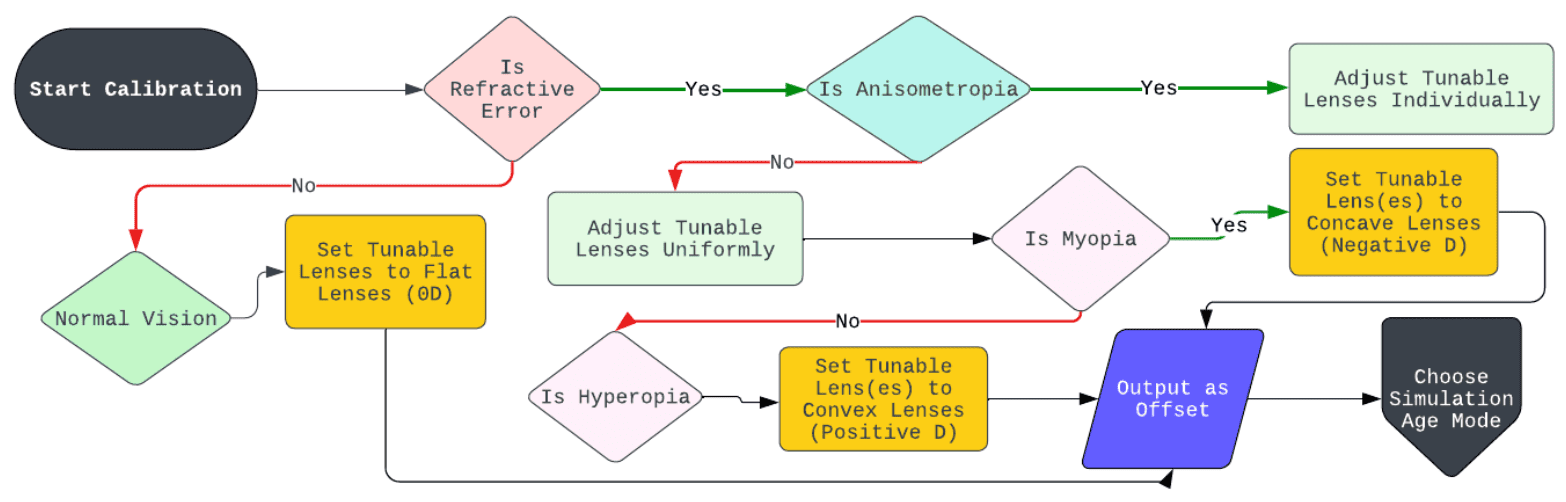}
    \caption{Calibration process for the tunable lenses presbyopia simulator. This flowchart outlines the step-by-step procedure for adjusting the tunable lenses to account for the wearer's existing refractive errors before initiating the presbyopia simulation.}
    \label{fig:calibrationFlowChart}
\end{figure*}

\end{document}